# CEPC DESIGN PERFORMANCE CONSIDERATIONS[1]

M. Koratzinos, University of Geneva and CERN, Geneva, Switzerland

*Abstract*

In this paper I will commend on the early CEPC design as of October 2014. In particular I will comment on the choice of circumference, minimum and maximum energy, number of collision points and target luminosity. I will finish with suggestions to increase performance with minimum incremental cost.

## CEPC DESIGN PHILOSOPHY

The design of the CEPC revolves around the philosophy of keeping costs low while achieving as much of the performance of an ultimate machine. For this reason the scope has been limited to primarily a Higgs factory, operating at a beam energy of 120GeV. Tunnel size has been kept to 54.8 kms, approximately twice as big as LEP. Two experiments are envisaged for $e^+e^-$ operation and a single beam pipe design has been chosen.

We will try to quantify the cost of these choices in terms of performance. In the late part of the paper we will also make suggestions to improve the performance of the baseline design.

## LUMINOSITY OF A CIRCULAR COLLIDER

The luminosity of a circular collider is given by

$$\mathcal{L} = \frac{3}{8\pi} \frac{e^4}{r_e^4} P_{tot} \frac{\rho}{E_0^3} \xi_y \frac{R_{hg}}{\beta_y^*} \quad (1)$$

where $r_e$ and $e$ are the classical radius of the electron and its charge, $P_{tot}$ the total SR power dissipated by one beam, $\rho$ the bending radius, $E_0$ the beam energy, $\xi_y$ vertical beam-beam parameter, $\beta_y^*$ the vertical beta function at the interaction point and $R_{hg}$ the geometric hourglass factor.

The maximum achievable $\xi_y$ depends on if a specific machine is beam-beam or beamstrahlung dominated [1].

### The beam-beam limit

The beam-beam limit depends on the damping decrement $\lambda_d$, the amount of energy loss when electrons move from one IP to the next:

$$\lambda_d = \left(\frac{U_0}{E}\right) \frac{1}{n_{IP}} \quad (2)$$

Where $U_0$ is the energy loss per electron in one turn. The LEP data has been used to derive this number following the formulation in [2]:

$$\xi_y^{max} \propto \lambda_d^{0.4} \quad (3)$$

which when fitted to the maximum beam-beam parameters achieved at LEP yields the approximate formula

$$\xi_y^{max} \approx 0.86 \cdot \lambda_d^{0.4} \quad (4)$$

We need to stress here that the above formulation is only based on a limited amount of LEP data and should be taken with a grain of salt. Beam-beam simulations and ultimately measurements on a real machine would provide a more accurate estimation, but for the purposes of this paper we consider the approach above adequate.

### The beamstrahlung limit

The beamstrahlung limit [3] is due to the fact that at high energies and luminosities beamstrahlung, the synchrotron radiation emitted by an incoming electron in the collective electromagnetic field of the opposite bunch at an interaction point, reduces beam lifetimes to values where the top-up injector cannot cope. The effect of beamstrahlung is very implementation specific and can be mitigated by small vertical emittance and large momentum acceptance.

Two analytical calculations exist for computing beam lifetimes due to beamstrahlung [3] [4] offering fast estimates of the effect. Analytical simulations assume Gaussian distributions (i.e. without non-Gaussian tails) and have other approximations. Therefore it is important to be checked against a complete simulation such as the one by K. Ohmi [5]. The comparison between the two analytical calculations and the simulation for two different energies (where beamstrahlung plays a crucial role in defining the beam lifetime) and for the specific implementation of FCC-ee [6] is shown in Figure 1 and Figure 2. Care is taken to use the effective $\beta_x^*$ coming out of the simulation, rather than the design value.

Both analytical calculations show reasonable agreement for momentum acceptances of interest here (between 1.5% and 2%) at beam energies of both 120GeV and 175GeV. This justifies the use of the analytic formulas instead of the much more accurate but time-consuming simulation for the purposes of this work.

The two regimes (the beam-beam dominated and the beamstrahlung dominated one), for the specific implementation of FCC in [6], can be seen in Figure 3. Such a machine would be beamstrahlung dominated at

---

[1] Talk title: Choice of circumference, minimum & maxim energy, number of collision points, and target luminosity

175GeV, but at lower energies will be beam-beam dominated.

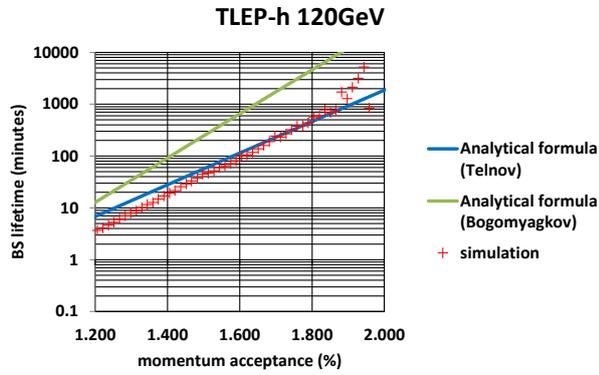

**Figure 1**: Beamstrahlung lifetimes for FCC-ee at 120GeV: comparison between the two analytical formulas and simulation.

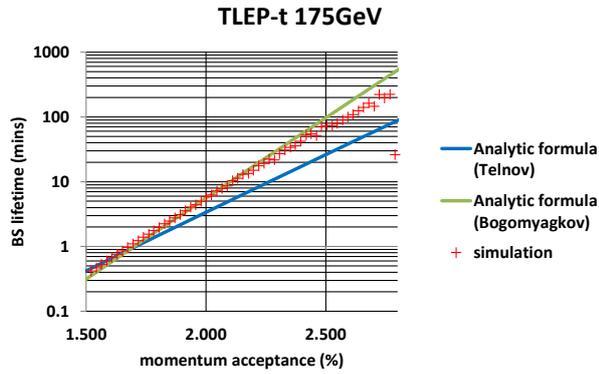

**Figure 2**: Beamstrahlung lifetimes for FCC-ee at 175GeV: comparison between the two analytical formulas and simulation.

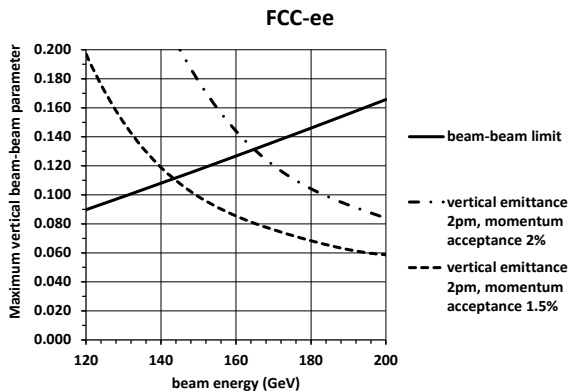

**Figure 3**: The beam-beam (solid line) and beamstrahlung limits for FCC-ee **[6]**. The beamstrahlung curves have been obtained assuming a beam lifetime of 300 seconds and two different momentum acceptance values using to the analytical formula in **[3]**. The machine is beamstrahlung dominated above approximately 165 GeV if the momentum acceptance is 2%.

The CEPC has different parameters, therefore the beamstrahlung curves are different. Regarding the beam-beam curve, the smaller diameter of the CEPC makes damping stronger and therefore a higher beam-beam parameter should be achievable in theory, however the CEPC approach uses a more conservative extrapolation of the maximum beam-beam parameter. As in the FCC-ee case, the machine is beam-beam limited at 120 GeV. The two regime plot of the CEPC can be seen in Figure 4.

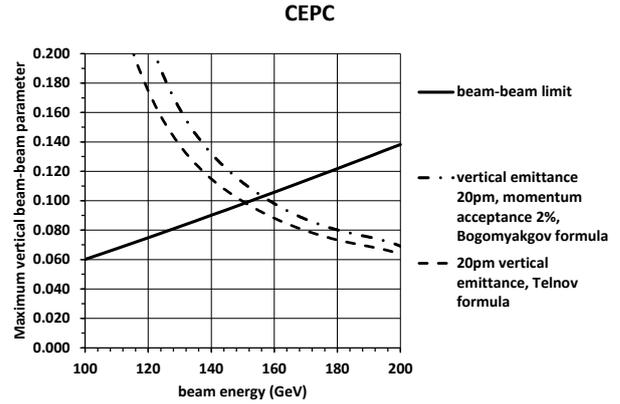

**Figure 4:** The beam-beam (solid line) and beamstrahlung limits for a collider with parameters of the CEPC. The beamstrahlung curves have been obtained assuming a beam lifetime of 300 seconds and a momentum acceptance and 2% for both analytical formulas in [4] and [3]. The machine is beamstrahlung dominated above approximately 155 GeV.

## RING CIRCUMFERENCE

As seen from eqn. (1), going to larger ring sizes increases the luminosity due to the increase in bending radius (linearly) but decreases it due to a smaller damping decrement (to the power 0.4, if we follow the parameterisation in [2]). The overall effect is an increase of luminosity with the ring circumference to the power 0.6. Therefore increasing the ring circumference to 70km ring increases luminosity by 20%. Going to a 100 km ring increases luminosity by 45%.

**Table 1:** Comparison of maximum luminosity for the CEPC design with the baseline tunnel circumference (53.6kms) and a larger ring (70kms). The beam-beam parameter decreases due to less damping for the bigger ring. Parameters affecting the bunch length $\sigma_z$ have been kept the same.

| Ring size | $\xi_y$ | $\sigma_z$ (mm) | Luminosity ($10^{34}$) |
|---|---|---|---|
| 53600m | .075 | 2.7 | 1.8 |
| 70000m | .067 | 2.7 | 2.2 |

The gain is modest, so the choice of tunnel size will also depend on other parameters, for example the desire to run at 175 GeV, which becomes easier with a larger ring, and the pp option later on, whose energy reach again benefits from a larger ring.

## NUMBER OF EXPERIMENTS

The CEPC design foresees two experiments. If four experiments were available, then the damping decrement between IPs would decrease by a factor of 2, and the maximum beam-beam parameter, following always the parameterisation in [2], would decrease by about 30%, as would the luminosity per experiment since we are at the beam-beam regime. Therefore having twice as many experiments would finally increase the luminosity yield per year by around 50%.

**Table 2:** Comparison of luminosity delivered to experiments as a function of the number of installed experiments. Damping decreases going from two to four experiments and therefore so does the corresponding maximum beam-beam parameter

| Energy | IPs | $\xi_y$ | Luminosity per IP ($10^{34}$) | Lumi*$N_{IP}$ ($10^{34}$) |
|---|---|---|---|---|
| 120GeV | 2 | 0.075 | 1.84 | 3.68 |
| 120GeV | 4 | 0.057 | 1.40 | 5.60 |

## RUNNING AT 45 GEV

45 GeV running is a highlight of the FCC-ee programme, as very high luminosities can be achieved. This comes at the expense of having many - O(10,000) - bunches circulating. Since the machine is beam-beam dominated at low energies, increasing the emittance in both planes decreases the number of bunches in the machine without affecting overall performance. In the case of FCC-ee emittances are increased by a factor 15 compared to 120GeV running. The CEPC design with a single beam-pipe and the pretzel scheme for bunch separation would not be able to run with more than O(100) bunches per species. This will severely limit the luminosity of the CEPC at the Z peak, but will also limit the synchrotron radiation power loss. The maximum beam-beam parameters can here be predicted with some confidence as the CEPC with two experiments and twice the size has the same damping decrement as LEP (=0.045). $\beta_x$ was adjusted to 0.4 m to give equal beam-beam parameters in x and y**.** Table 3 shows the essential parameters of the CEPC running at 45 GeV. The first row is simply extrapolating from the 120GeV design keeping emittances the same, the second is the result of increasing emittances by a factor of 20 in an attempt to limit the number of bunches and the third is limiting also the power to 20% of the initial power. The resulting number of bunches – 160 – might just be manageable with a single beam-pipe and the delivered luminosity would be a factor of 14 smaller than the advertised performance of FCC-ee at 45GeV and four experiments. Finding a way to accommodate more bunches would increase the performance at 45 GeV dramatically.

**Table 3**: Synchrotron radiation power per beam, number of bunches and expected luminosity for different schemes for CEPC 45GeV running

| Scheme | Power (MW) | No. of bunches | Lumi ($10^{34}$) |
|---|---|---|---|
| **Extrapolating from 120GeV** | 50 | 15500 | 27 |
| Increasing emittances by 20 | 50 | 780 | 21 |
| **Reducing power by 5** | 10 | 160 | 4 |

## RUNNING AT 175 GEV

Running above the $t\bar{t}$ threshold is considered an important part of the physics programme of FCC-ee but is not a high priority for the CEPC. The smaller ring diameter would make the energy loss per turn due to synchrotron radiation higher, needing more RF voltage. Both the CEPC and FCC-ee would be running in a beamstrahlung-dominated regime at this energy. The energy loss per turn for the CEPC is 13.6GeV. This necessitates an RF system twice as big as that of FCC-ee, with around 1200m of RF cavities. The performance of both machines can be compared in Table 4. There is a performance loss of a factor 5 compared to the parameters of FCC-ee.

**Table 4**: Energy loss per turn, emittances, number of bunches and projected luminosities for 175GeV running for the CEPC and FCC-ee circular colliders

| Ring | E_loss (GeV) | $\epsilon_x/\epsilon_y$ (nm/pm) | No. bunches | Luminosity ($10^{34}$) |
|---|---|---|---|---|
| CEPC | 13.6 | 7/10 | 7 | 0.8 |
| FCC-ee | 7.6 | 2/2 | 98 | 1.8 |

## SUGGESTIONS FOR IMPROVED PERFORMANCE

Here we examine if there are avenues to pursue to increase the performance of the CEPC while sticking on the strategic choices of ring diameter, number of experiments and single beam-pipe design.

Effectively, the largest limitation comes from the constrained number of bunches due to the single beam-pipe design. The disadvantage becomes more pronounced if one is to run at lower energies and more specifically at 45GeV, which adds an extra dimension to the physics case of the machine. Another disadvantage is that a crab waist scheme that looks promising for the FCC-ee (admittedly for energies lower than 120GeV) cannot be implemented at the CEPC with the single beam-pipe option.

There is a way, however, to mostly keep the single beam-pipe design philosophy and at the same time allow for more bunches: the introduction of a double beam pipe in two or four straight sections and the introduction of a bunch train scheme. Electrostatic separators can be used to separate the beams at the edges of the relevant straight sections initially and magnetic elements can take over after the beams are separated sufficiently. In such a scheme, one or two bunch trains of electrons and positrons can circulate with no parasitic collisions, provided that the total length of the bunch trains is less than the length of the double-pipe straight sections. In this scheme and with a total length of the double pipe straight sections of 4 kms, 2000 electron and positron bunches can be accommodated with a 2 m (7 nsec) separation. The increase in cost is modest, as less than 10% of the machine is equipped with a double beam pipe. The arcs keep the single beam-pipe design. Such a design can naturally accommodate a crab waist collision scheme.

Another weakness of the design is the inconsistent bunch length compared to the $\beta_y^*$ value (2.7mm compared to 1.2mm). $\beta_y^*$ has correctly been set to a low value, as it directly affects luminosity. Here the value chosen is 20% larger than that of the FCC-ee design. However, the fact that the bunch length is 2.3 times the $\beta_y^*$ value introduces a large geometrical loss factor (hourglass value of 0.68) and, more importantly, gives rise to beam instabilities that result in lower luminosities. On the other hand, the large bunch length masks the fact that the design is in reality beamstrahlung limited, had the bunch length been as short as in the FCC-ee design – indeed, in such a case the beam lifetime drops to a bit more than a minute.

Looking closely at the parameters, one realises that the CEPC design has a factor of 7/10 larger horizontal/ vertical emittance compared to the FCC-ee design. The emittance ratio of 330 used in the CEPC design is a realistic number and close to what was achieved at LEP (250) – the FCC-ee design pushes this number to 500 or even 1000 for the $t\bar{t}$ running. However, the horizontal emittance of the CEPC design is too large when seen in context of what the FCC-ee design expects to achieve. The ways to reduce the horizontal (and, therefore vertical) emittance include:

- The introduction of stronger focusing by adopting $90^o$ FODO optics instead of the current $60^o$ CEPC design for the horizontal plane. Theoretically, going from $60^o$ to $90^o$ optics reduces the emittance by a factor 3.
- Adopt a shorter FODO cell. The bending angle of the CEPC is twice as big as the FCC-ee one per unit length, and the emittance depends on the third power of the bending angle. Going to a FODO length of 38 m from the current 47.2 m would give a factor of 2 lower emittance.

LEP run in various configurations, but 90/60 optics worked well.

It should be mentioned here that above mentioned solutions have repercussions regarding the cost of the magnet system of the machine.

The momentum compaction factor could also be reduced (the FCC-ee design has a momentum compaction factor which is a factor of 7 smaller) and at least a factor of 2 reduction will be necessary to avoid having a bunch length much larger than $\beta_y^*$. Moving from a phase advance of $60^o$ to $90^o$ would reduce the momentum compaction factor by a factor 2. Reducing the FODO length to 38 m would give another factor 1.5, for a total factor 3 reduction.

This suggestion for the parameter set of CEPC can be seen in Table 5 (modifications in bold typeface). Essentially the horizontal and vertical emittances have been reduced by a factor 3 (this should be achievable by only going to a $90^o$ optics without any shortening of the FODO length). The momentum compaction factor has also been reduced by 3. This can be achieved by a combination of $90^o$ optics and shorter FODO length. The CEPC philosophy of a higher beam-beam parameter in the horizontal plane has been followed, although the CERN design uses similar beam-beam parameters in both planes. Due to the lower momentum compaction number the bunch length is now reduced to acceptable values (the equilibrium bunch length is calculated to be 1.55mm). Since the horizontal and vertical beam size is now reduced due to the smaller emittances and the bunch length is also reduced, less charge per bunch is called for to avoid beamstrahlung lifetime problems. This increases the total number of bunches to 120. This is still possible (but marginal perhaps) with the pretzel (single beam-pipe) scheme, but is a non-issue in the bunch separation scheme proposed above. The luminosity shows a healthy increase to above $3 \times 10^{34} cm^{-2} s^{-1}$.

## ACKNOWLEDGMENTS

This paper would not be possible without the valuable input of many colleagues, I would like to single out Dmitry Shatilov for his helpful insight to the problems discussed here.

## REFERENCES

[1] M. Koratzinos, «Performance limitations of circular colliders: head-on collisions,» *CERN-ACC-NOTE-2014-0066*.

[2] R. Assmann and K. Cornelis, «The beam-beam limit in the presence of strong synchrotron radiation damping,» *CERN-SL-2000_046 OP*.

[3] A. Bogomyagkov et al., «Beam-beam effects investigation and parameters optimization for a circular e + e − collider at very high energies,» *Phys.Rev.ST Accel.Beams 17 (2014) 041004*.

[4] V. Telnov, «Restriction on the energy and luminosity of e+e- storage rings due to beamstrahlung,» *Phys. Rev. Letters 110, 114801 (2013) arXiv:1203.6563*.

[5] K. Ohmi, F. Zimmermann, «FCC-ee/CepC Beam-Beam Simulations with Beamstrahlung,» chez *Proceedings of IPAC2014, Dresden, Germany*, 2014.

[6] Wenninger, J. et al., «Future Circular Collider Study Lepton Collider Parameters,» *CERN EDMS no. 1346082,* 2014.

**Table 5:** The CEPC parameter set as was presented on 10 October 2014 and as suggested in this paper.

| Parameter | Unit | Value 10/10/2014 | Value – this suggestion |
|---|---|---|---|
| Beam energy [E] | GeV | 120 | 120 |
| Circumference [C] | m | 54752 | 54752 |
| Number of IPs [$N_{IP}$] | | 2 | 2 |
| SR loss/turn [$U_0$] | GeV | 3.11 | 3.11 |
| Bunch number/beam [$n_B$] | | 50 | **120** |
| Bunch population [$N_e$] | | 3.79E+11 | **1.5E+11** |
| SR power/beam [P] | MW | 51.7 | **50** |
| Beam current [I] | mA | 16.6 | 16.6 |
| Bending radius [$\rho$] | m | 6094 | 6094 |
| momentum compaction factor [$\alpha_p$] | | 3.36E-05 | **1.1E-05** |
| Revolution period [$T_0$] | s | 1.83E-04 | 1.83E-04 |
| Revolution frequency [$f_0$] | Hz | 5475.46 | 5475.46 |
| emittance (x/y) | nm | 6.12/0.018 | **2/0.006** |
| $\beta^*$(x/y) | mm | 800/1.2 | 800/1.2 |
| Transverse beam size (x/y) | μm | 69.97/0.15 | **40/0.085** |
| $\xi_x/\xi_y$ per IP | | 0.118/0.083 | **0.146/0.104** |
| Bunch length SR [$\sigma_{s.SR}$] | mm | 2.14 | **1.24** |
| Bunch length total [$\sigma_{s.tot}$] | mm | 2.65 | **1.55** |
| Lifetime due to Beamstrahlung | min | 47(simulation) | **68/360 (analytical)** |
| lifetime radiative Bhabha scattering [$\tau_L$] | min | 51 | **34** |
| RF voltage [$V_{rf}$] | GV | 6.87 | 6.87 |
| RF frequency [$f_{rf}$] | MHz | 650 | 650 |
| Harmonic number [h] | | 118800 | 118800 |
| Synchrotron oscillation tune [$\nu_s$] | | 0.18 | **0.10** |
| Damping partition number [$J\varepsilon$] | | 2 | 2 |
| Energy spread SR [$\sigma_{\delta.SR}$] | % | 0.132 | 0.132 |
| Energy spread BS [$\sigma_{\delta.BS}$] | % | 0.096 | **0.099** |
| Energy spread total [$\sigma_{\delta.tot}$] | % | 0.163 | **0.165** |
| $n_\gamma$ | | 0.23 | **0.17** |
| Transverse damping time [$n_x$] | turns | 78 | 78 |
| Longitudinal damping time [$n_\varepsilon$] | turns | 39 | 39 |
| Hourglass factor | | 0.68 | **0.81** |
| Luminosity /IP | cm$^{-2}$s$^{-1}$ | 2.04E+34 | **3.07E+34** |
| FODO length | m | 48 | **38** |
| FODO phase advance (horiz./vertical) | degrees | 60/60 | **90**/60 |